\documentclass[twocolumn,prl,showpacs,floatfix,superscriptaddress]{revtex4-1}%
\usepackage{graphicx}
\usepackage{dcolumn}
\usepackage{bm}
\usepackage{amssymb}
\usepackage{amsmath}
\usepackage{amsfonts}
\usepackage{xcolor}%
\usepackage{float}
\definecolor{darkblue}{rgb}{0.1,0,0.5}
\usepackage[colorlinks,hyperindex]{hyperref}
	\hypersetup
	{
		colorlinks,%
		citecolor=darkblue,%
		linkcolor=darkblue,%
		urlcolor=darkblue,%
	}
\begin{document}
\title{Streaking temporal double slit interference by an orthogonal two-color laser field}

\author{Martin Richter}
\email{richter@atom.uni-frankfurt.de}
\affiliation{Institut f\"ur Kernphysik, Goethe-Universit\"at Frankfurt, 60438
Frankfurt am Main, Germany}

\author{Maksim Kunitski}
\affiliation{Institut f\"ur Kernphysik, Goethe-Universit\"at Frankfurt, 60438
Frankfurt am Main, Germany}

\author{Markus Sch\"offler}
\affiliation{Institut f\"ur Kernphysik, Goethe-Universit\"at Frankfurt, 60438
Frankfurt am Main, Germany}

\author{Till Jahnke}
\affiliation{Institut f\"ur Kernphysik, Goethe-Universit\"at Frankfurt, 60438
Frankfurt am Main, Germany}

\author{Lothar P.H. Schmidt}
\affiliation{Institut f\"ur Kernphysik, Goethe-Universit\"at Frankfurt, 60438
Frankfurt am Main, Germany}

\author{Min Li}
\email{minli@pku.edu.cn}
\affiliation{State Key Laboratory for Mesoscopic Physics and Department of Physics, Peking University, Beijing 100871, People's Republic of China}

\author{Yunquan Liu}
\affiliation{State Key Laboratory for Mesoscopic Physics and Department of Physics, Peking University, Beijing 100871, People's Republic of China}
\affiliation{Collaborative Innovation Center of Quantum Matter, Beijing 100871, China}

\author{Reinhard D\"orner}
\affiliation{Institut f\"ur Kernphysik, Goethe-Universit\"at Frankfurt, 60438
Frankfurt am Main, Germany}
\date{\today}

\begin{abstract}
We investigate electron momentum distributions from single ionization of Ar by two orthogonally polarized laser pulses of different color. The two-color scheme is used to experimentally control the interference between electron wave packets released at different times within one laser cycle. This intracycle interference pattern is typically hard to resolve in an experiment. With the two-color control scheme these features become the dominant contribution to the electron momentum distribution. Furthermore the second color can be used for streaking of the otherwise interfering wave packets establishing a which-way marker. Our investigation shows that the visibility of the interference fringes depends on the degree of the which-way information determined by the controllable phase between the two pulses.

\end{abstract}

\maketitle



Electron wave packets launched from a sample at different positions \cite{Cohen66,Akoury07science,Lai13pra} or at different times \cite{paulus05prl,Voitkiv11prl} give rise to interference effects in the final electron momentum distribution whenever the wave packets cannot be distinguished by a measurement. Any which-way information will destroy the interference. There are at least two ways to record such which-way information. One is to store it in another particle by entanglement of the electron with that particle \cite{Schoeffler08science,Akoury07science} or the environment \cite{Sonnentag07prl,arndt99,Duerr98PRL}. The second is by marking the which-way information in the electron itself, either in a spin degree of freedom \cite{Summhammer82PhysLettA} or in a motional degree of freedom like one of the momentum components. 

A versatile scenario to create electron wave packets for which several prominent interference effects have been identified in recent years is strong-field ionization in an ultrashort laser pulse. There the wave packets are released by strong-field tunnel ionization during the laser pulse and are subsequently driven by this optical field (see Fig. 1a). The interference fringes are then observed in the final electron momentum distribution long after the laser pulse. In the present work we show how one of these strong-field ionization interference patterns can be made visible and switched on and off. We have achieved this by using a second, phase locked orthogonal laser field of doubled frequency which encodes the which-way information in one momentum component.

\begin{figure}[h]
\centering
\includegraphics[width=0.9\columnwidth]{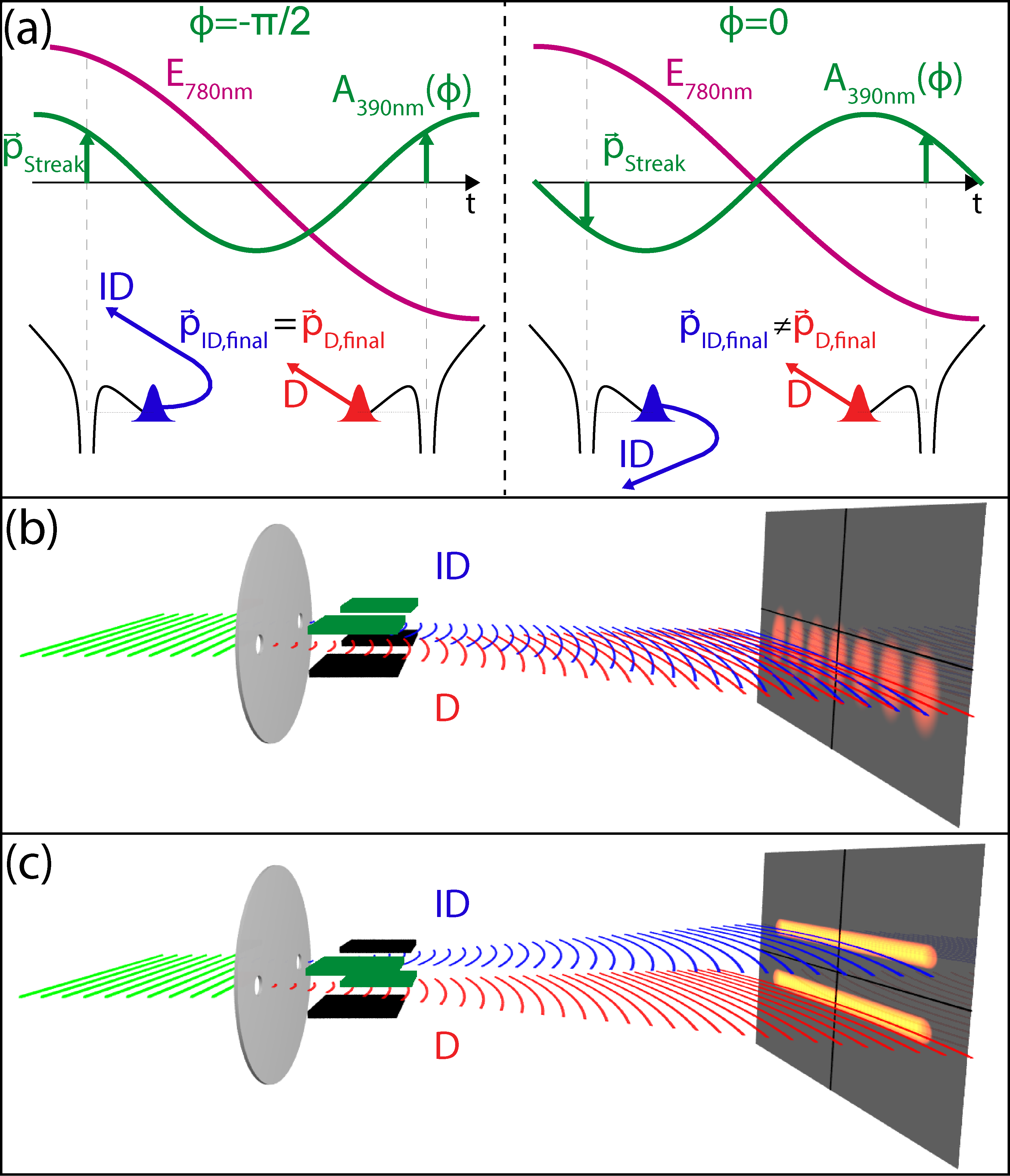}
\caption{(a) The vector potential of the second harmonic field relative to the ionizing 780 nm electric field and the electron trajectories ID (indirect) and D (direct) at two different phases between the colors (see eq. 1). Though the fields are perpendicularly polarized in the experiment, here they are drawn in parallel for a more intuitive understanding of the streaking dynamics. For $\phi=-\pi/2$ both trajectories are streaked in the same direction allowing them to interfere in momentum space. For the phase $\phi=0$ trajectories A and B are streaked to opposite directions which extinguishes the interference. (b) is a sketched spatial analogue where a double slit (depicted with two holes in a disk) transforms an incoming plane wave into two coherent spherical waves (for the sake of simplicity shown with curved lines). These waves are then steered into the same direction by two plane deflectors (capacitors) of equal polarity resulting in an interference pattern. This corresponds to the phase $\phi=-\pi/2$. (c) is the case for $\phi=0$ where the two waves are deflected in opposite directions showing no interference.}
\label{intro}
\end{figure}

The most prominent interference fringes emerging in strong-field ionization are the equidistant peaks in the electron energy distribution. These above threshold ionization (ATI) peaks are spaced by the photon energy. They result from the interference of wave packets born periodically in time at subsequent laser cycles. In experiments at higher laser intensities these structures are washed out by the averaging over the different intensities in the focus since the energy offset of this comb depends on the laser intensity. Much less striking and only discovered in 2005 \cite{paulus05prl} are additional fringes resulting from the interference between wave packets born within one cycle at times where the vector potential is the same, but the direction of the electric field is opposite (trajectories ID and D in Fig. 1a). We refer to this channel in the following as intracycle interference. These interferences have been seen first in an experiment using three-cycle (6 fs, 760 nm) laser pulses \cite{paulus05prl}. In a later experiment two-color pulses have been used to characterize the phase difference between the two wave packets after tunneling \cite{kitzler2012prl}. Most of the works discussing the intracycle interferences are theoretical \cite{burgdoerfer10pra,burgdoerfer10pra2,bauer12pra}. The main reason is that in experiments with linearly polarized multi-cycle pulses these interference structures are buried in a wealth of other, more prominent structures. They show up only as a height modulation of the ATI peaks \cite{kitzler2012prl,burgdoerfer14pra}. 
ATI and intracycle interferences would occur even without the influence of the ionic potential. The unavoidable presence of this potential gives rise to further structures in the momentum distribution of electrons upon strong-field ionization, which also obscure the intracycle interference fringes. The key physical effect behind these additional structures is that electron trajectories (labeled with ID in Fig. 1a) which escape in one direction, are turned around by the oscillating laser field and they are deflected by passing the ionic core. This deflection gives rise to what has been named "Coulomb focusing" \cite{corkum96pra} leading to a narrowing of the momentum distribution perpendicular to the field direction \cite{staudte13prl}. They also lead to spider-leg shaped structures \cite{vrakking11science,vrakking12prl,vrakking12prl2} (labeled with S in Fig. 2a). 

In the present work we show that orthogonally polarized two-color pulses (OTC) \cite{kitzler05prl,staudtekitzlerprl14,kitzler14pra} can be used to turn the typically faint intracycle interference fringes into a dominating structure in the electron momentum distribution and at the same time can be used as a controllable which-way marker allowing to efface the interferences. The OTC pulses are shown in Fig. 1a. We use a strong 780 nm pulse ($1.4 \cdot 10^{14} \text{W/cm}^{2}$) and a weak ($1.3 \cdot 10^{13}  \text{W/cm}^2$) 390 nm pulse. The conditions are chosen such that the tunneling is mainly caused by the 780 nm pulse while the orthogonal 390 nm field mildly streaks the electron wave packet. By changing the phase between the two colors we can adjust the vector potential of the 390 nm laser field which causes the streaking such that it is the same for trajectories ID and D (Fig. 1a, left). In this case ID and D are indistinguishable, there is no which-way information and we expect the intracycle interference to occur. Alternatively, the phase between the 390 and 780 nm laser pulses can be chosen such that the vector potential of the 390 nm laser field is opposite at points ID and D (Fig. 1a, right). In this case the 390 nm laser field marks the slits in time and makes the wave packets distinguishable switching off the interference. 

A spatial analogue of this scenario is shown in Figs. 1b and 1c. An electron wave traverses a double slit where behind each slit a pair of deflector plates is mounted. The deflection is orthogonal to the  interference fringes. If the deflectors behind both slits are biased with the same polarity they both deflect the electron wave packet to the same direction and an interference pattern occurs. If however, the polarity is opposite, one deflects upward, one downward. The which-slit information is then imprinted in the  momentum component orthogonal to the interference fringes and no double slit interference occurs.


In the experiment the OTC field
\begin{align} \label{eq-otc-feld}
\vec{E}=E_{z, 780}  \cos(\omega t) \vec{e}_z +E_{y, 390}  \cos(2 \omega t + \phi) \vec{e}_y
\end{align}
is used, where $\phi$ is the tunable phase between the two colors. The three-dimensional electron momenta were measured in coincidence with argon ions using COLTRIMS \cite{Doerner200095,Jagutzki02nuc}. Further experimental details are given in \cite{supplemental}.


\begin{figure}[h]
\centering
\includegraphics[width=0.9\columnwidth]{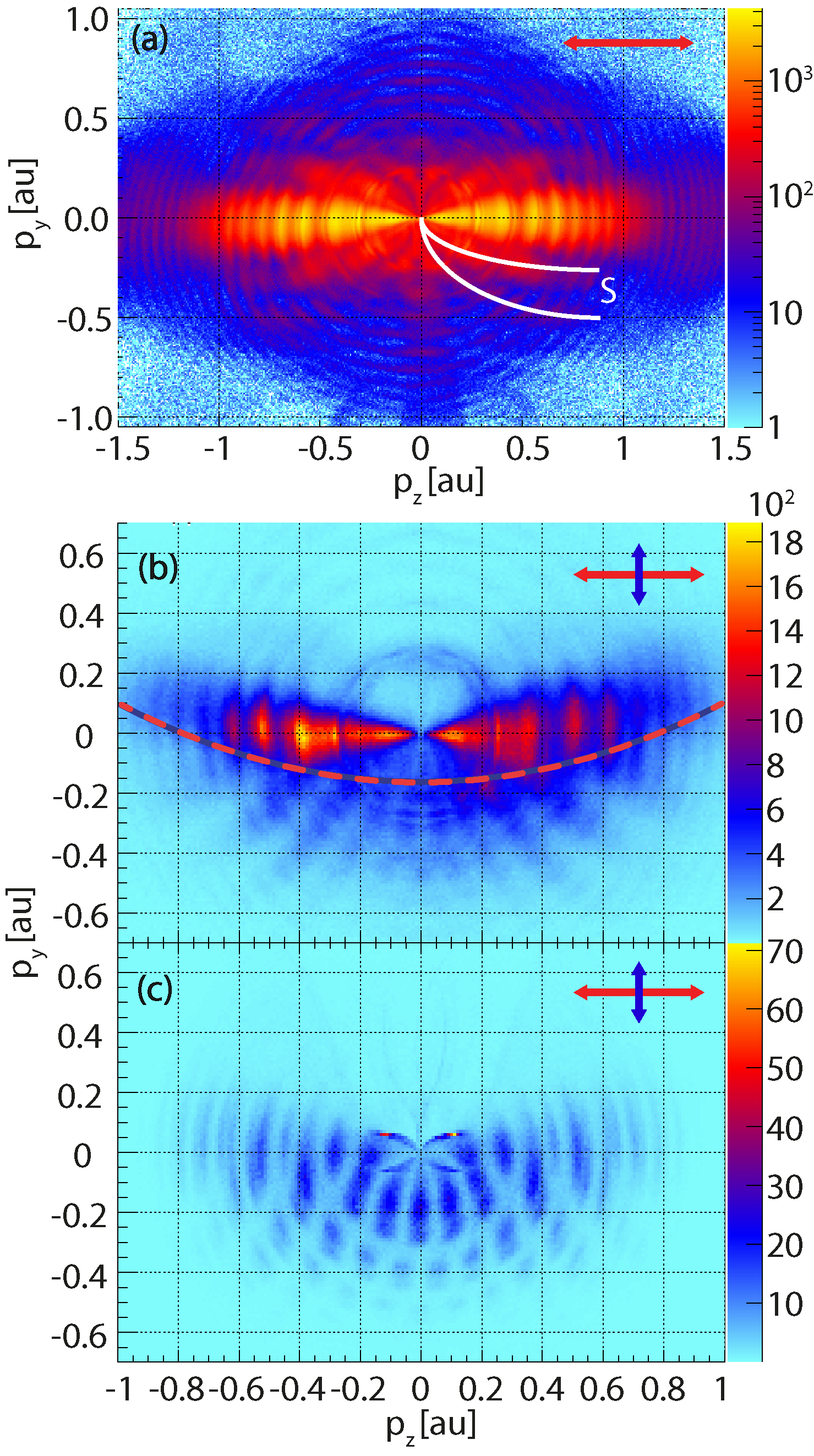}
\caption{Electron momentum distribution from strong-field single ionization of Argon. The data are integrated over an angular range $\vartheta=90\pm15 ^\circ$ where $\vartheta = \text{acos}(p_x/\sqrt{p_x^2+p_y^2+p_z^2})$ is the angle between the electron momentum vector and the normal to the $(p_z,p_y)$ plane. (a) Experiment with 780 nm ($1.4 \cdot 10^{14} \text{W/cm}^2$, 40 fs) pulse only. The laser polarization direction is shown by an arrow. (b) 780 nm/390 nm orthogonal two-color pulse with an intensity ratio $I_{390}/I_{780}=0.09$ and a phase difference $\phi=-\pi/2$. The polarizations of the 780 nm and 390 nm lights are shown by red and blue arrows, respectively. The field driven momentum $p=-A(t_0)$ is shown by the dashed line. The finger-like structure results from the intracycle interference. (c) QMTC calculation for same laser parameters as in (b).}
\label{ex_qtmc}
\end{figure}

The momentum distribution of electrons originating from single ionization of Argon by a single-color 780 nm pulse with an intensity of $1.4\cdot 10^{14}  \text{W/cm}^2$ is shown in Fig. 2a. The distribution exhibits the features which are well known from the literature \cite{Xie12prl,staudte13prl,vrakking11science}. Namely, it shows a cutoff at a momentum of approx. $p_z^{cutoff}=1.08$ au which is the maximum momentum $p_z$ an electron can acquire in the 780 nm field at this intensity without rescattering at the nucleus. A decrease of intensity at this momentum is visible in Fig. 2a (note the logarithmic color scale). Electrons at larger momenta than $p_z^{cutoff}$ originate from backscattering at the nucleus and form what is known as the ``plateau'' in the energy spectrum \cite{Paulus94JPhyB}. The ATI peaks are visible as rings. In the regime of energies below $2 {U_P}$ (momenta below $p_z^{cutoff}$) the dominating emission is along the polarization axis with small transverse momenta. This feature is caused by the Coulomb focusing of electrons which pass the nucleus (e.g. trajectory ID in Fig. 1a). Also the spider-leg shape holographic interference features can be seen (feature marked with S) \cite{vrakking11science,vrakking12prl,vrakking12prl2}. The intracycle interferences, however, are not visible in this figure without a detailed analysis. They are buried below the other, much more prominent structures. 

By adding a weak 390 nm streaking field orthogonal to the 780 nm field with a phase shift of $\phi=-\pi/2$, a strong finger-like structure appears in the lower half of the graph (Fig. 2b) in which the field driven momentum is represented by the dashed line. For these experimental parameters we performed a quantum-trajectory Monte Carlo (QTMC) simulation shown in panel (c). This simulation describes the strong-field ionization semiclassically by combining Ammosov-Delone-Krainov (ADK) theory and Feynman's path integral approach (see \cite{liu14prl} for details). In ADK-theory the ionization rate, the tunnel exit and the momentum distribution are prescribed \cite{Delone91JOSA}. During the laser pulse electron trajectories are launched with a probability and a transverse momentum distribution given by ADK-theory. These electrons are propagated classically in the laser field and the Coulomb field of the ionic core and the action integral along the trajectory is calculated for each electron. Using this phase information the contributions from different trajectories can be added coherently. Experiment and theory both show the finger-like lines which, in contrast to the spider-leg structure, do not end at $(p_z,p_y)=(0,0)$. The second harmonic streaking field reveals these interference structures which are otherwise hidden behind the dominating Coulomb focused trajectories. The QTMC calculation does not reproduce the events along $p_y=0$ visible in the experiment. This might be due to the fact that in the experiment the spatial and temporal overlap of the 780 nm and 390 nm pulse is never as perfect as assumed in the calculation (i.e. due to imperfect beam profiles). The intensity in the calculation is averaged over the focal volume. This realistic focal averaging is essential for visual comparison with the experiment to reduce the otherwise dominating contribution of the ATI peaks.      

\begin{figure}[h]
\centering
\includegraphics[width=0.9\columnwidth]{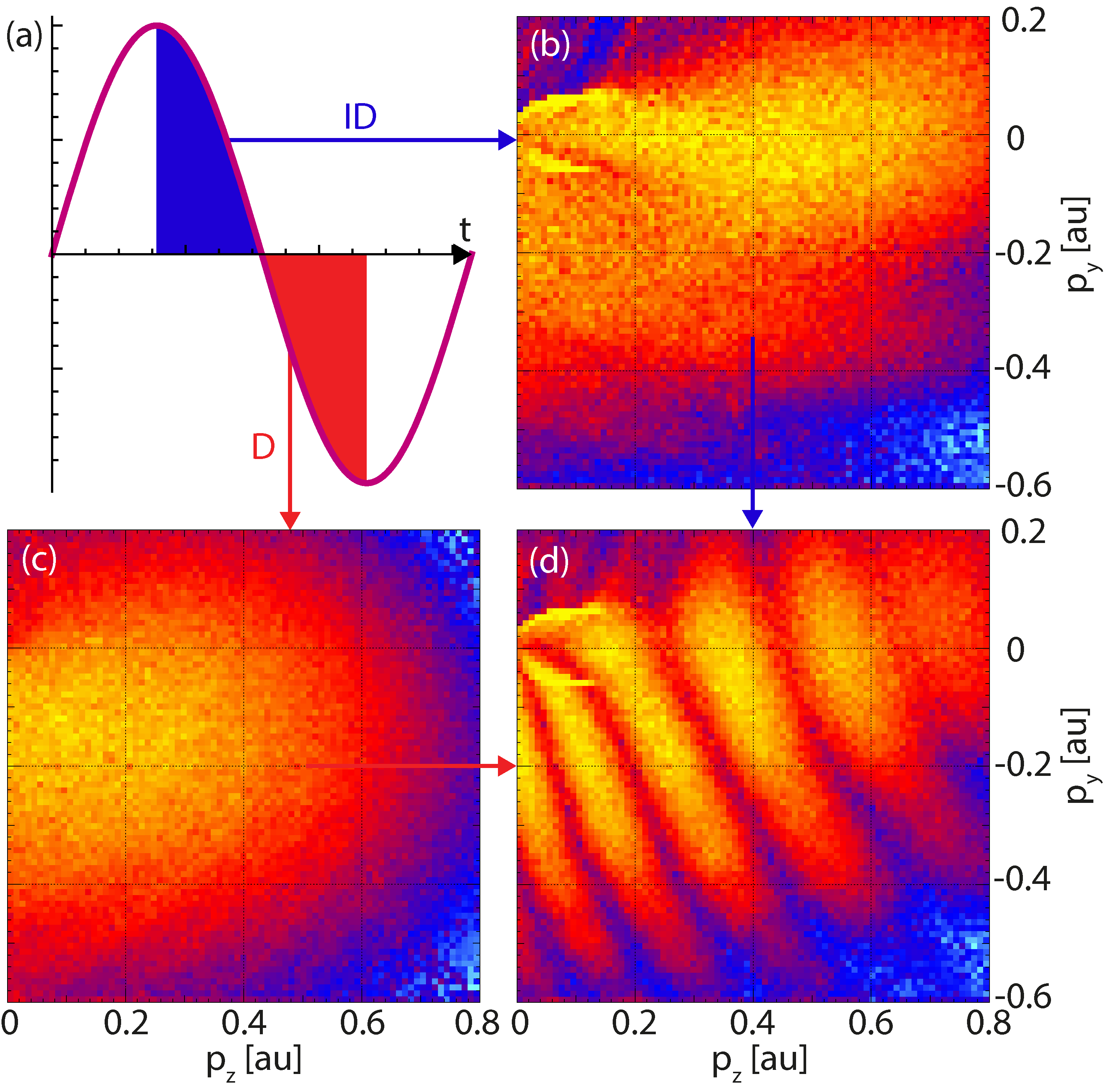}
\caption{In (a) the time windows are shown where the electron trajectories ID (blue) and D (red) are born in the electric 780 nm field. (b)-(d) represent the QTMC-calculated momentum spectra for these trajectories at a streaking field phase of $\phi=-\pi/2$. In (b) only the electrons tunneling during the blue marked quarter cycle are plotted. These indirect trajectories are driven back to the core by the laser field but only weakly interact with it because of streaking. The direct non-returning trajectories originate from the red marked quarter cycle and are shown in (c). Adding up the trajectories from both quarter cycles coherently (d) leads to the same intracycle interference pattern as in the experiment.}
\label{intid}
\end{figure}
 
With the help of the QTMC calculation we show that the new finger-like structures are indeed intracycle interference fringes (Fig. 3). We separate the trajectories in those starting in the quarter cycles marked in blue and red. The blue part of the trajectories are driven back by the 780 nm field and pass nearby the ion. The deflection of these trajectories leads to a partial focusing as visible in panel b. The trajectories from the red quarter cycle escape directly without recollision. They lead to a final momentum which is defined by the vector potential at the instant of ionization and is broadened by the initial transverse momentum distribution after tunneling. Separately, none of the two distributions from the quarter cycles show any finger-like structure. Adding the trajectories from both coherently, i.e. allowing for interference between wave packets from both quarter cycles (Fig. 3d), yields the finger-like structure which is visible in the experiment.

\begin{figure}[h]
\centering
\includegraphics[width=1.0\columnwidth]{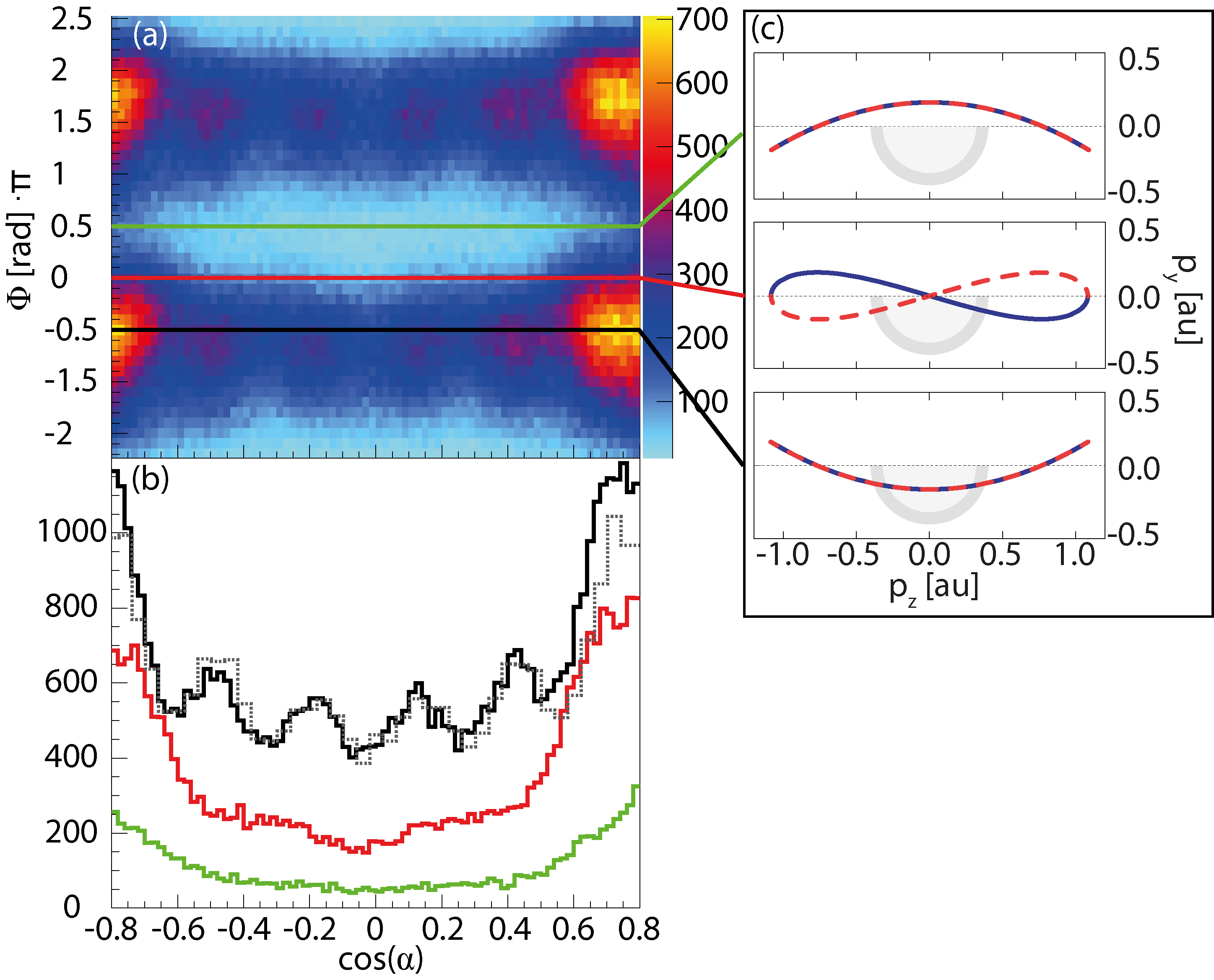}
\caption{Dependence of the intracycle interferences on the phase $\phi$ between the fields of the two colors (see eq. 1). (a) $\text{cos} (\alpha)=p_z/p_{total}$ with $p_{total}=\sqrt{p_x^2+p_y^2+p_z^2}$ is plotted against the phase $\phi$. To improve the interference contrast there is an additional restriction in momentum $0.32 \ \text{au}<p_{total}<0.4 \ \text{au}$ and $p_y<0$ which gives a half spherical shell in momentum space (shown in (c) in grey) and again as in Fig. 2 the data is integrated over an angular range $\vartheta=90\pm15 ^\circ$. Three projections along the $\text{cos} (\alpha)$-axis for the phases $-\pi/2$ (black), $0$ (red) and $\pi/2$ (green) are shown in (b). The grey dashed curve is from our QTMC calculation at $-\pi/2$ and agrees very well with the experiment. In (c) the OTC field's negative vector potentials corresponding to the field driven final momenta are shown. The blue and red curves correspond respectively to the quarter cycle time windows where trajectories ID and D originate (see Fig. 3).}
\label{cosphi}
\end{figure}

The two-color calculations and experimental data discussed so far where for a phase of $\phi=-\pi/2$ between the 390 nm and 780 nm field where the intracycle interference fringes were best visible. In Fig. 4 we investigate how the visibility of the intracycle fringes changes with the relative phase between the two colors. This phase is plotted on the vertical axis of Fig. 4a. To examine the interference we plot the events located along a half spherical shell in momentum space which cuts through the fingers. It is indicated in Fig. 4c by the gray shaded area. The fringe visibility changes strongly with the relative phase $\phi$. For $\phi=-\pi/2$ the black line in Fig. 4b shows the interference nicely, the gray histogram shows our QTMC calculation in excellent agreement with the experiment. $\phi=-\pi/2$ corresponds to the case where the 390 nm field steers the two trajectories ID and D to the same direction which leads to the intracycle interference. This can also be seen in Fig. 4c where the sum of the 390 nm and 780 nm light vector potential is shown. The blue curve indicates the first (ID), the red dashed curve the second quarter cycle (D). This corresponds to the situation where the deflectors behind the slits in Fig. 1b are biased with the same polarity. There is no which-way information imprinted by the 390 nm field and the wave packets show the maximum interference contrast. The contrast is not 100 $\%$ because the wave packet ID from the blue quarter cycle in Fig. 3b experiences some Coulomb focusing which drags some flux out of the grey circular region while the wave packet D from the red quarter cycle does not experience that loss. So the contributions from the two quarter cycles are different in amplitude and cannot lead to a complete destructive interference. The opposite scenario is given for $\phi=0$ shown by the red curve in Fig. 4b. Here the total flux is about a half of that at $\phi=-\pi/2$ but the interference has almost completely disappeared. In this case the vector potential of the 390 nm light has opposite signs for the two quarter cycles. This marks the slit, analogous to putting opposite polarity on the two deflectors in Fig. 1c. The corresponding sum of both vector potentials shown in Fig. 4c illustrates this. The contributions from the blue and the dashed red part of the laser cycle do not overlap and hence all quarter cycles have become distinguishable. As a consequence this switches off the intracycle interference. Finally for $\phi=\pi/2$ most of the flux in the lower half plane is gone. In this case the interference is in the upper half plane which is not visible in our graph.

In conclusion, controlling the laser field in two spatial dimensions with different frequencies gives full control over continuum electron wave packets from strong-field ionization. Previous works using this handle to modify higher harmonic generation \cite{marangosprl14} and non-sequential double ionization \cite{staudtekitzlerprl14} relied on the modification of the flux. In the present work we have shown that such control becomes even more valuable if one considers the phase of the wave packets. Since many of the observables of strong-field effects are caused by interference between different electron wave packets this method is very powerful, as we have proven by switching one of the interference effects on and off. A future application might involve using the interferogram for an easier extraction of the phase between the tunneling wave packets \cite{kitzler2012prl} for molecular orbital imaging. We have also shown that such control over interferences can be achieved by using an additional streaking field as a which-way marker. A similar scheme can be used for any of the structures in the electron emission in strong fields. One can envision that such switchable interferences can also be used for other important effects caused by the electron wave packets such as e.g. selective bond breakage in molecules \cite{Xie12prl}. For instance, by changing the phase between the OTC pulses one can select the energy of the recolliding electrons. In turn, this might be used for a selective population of doubly charged repulsive states upon recollision, favoring particular fragmentation channels.

\acknowledgments  
This work was funded by the Deutsche Forschungsgemeinschaft.
Y. L. acknowledges the supports by the 973 programs (No. 2013CB922403) and the NSFC of China (No.11125416, 11121091, 11434002).

\bibliographystyle{apsrev4-1}
\bibliography{completebib}






\end{document}